\documentclass{ws-procs9x6}

\begin{document}

\title{Five golden rules for superstring phenomenology}


\author{H. P. Nilles}

\address{Physikalisches Institut\\
Universit\"at Bonn\\ 
Nussallee 12\\
D-53115 Bonn, Germany\\ 
E-mail: nilles@th.physik.uni-bonn.de}



\maketitle

\abstracts
{Recent measurements of the values of gauge coupling constants
as well as neutrino properties support the idea of a grand unified (GUT)
description of particle physics at a large scale of 
$M_{GUT}\sim 10^{16}$ GeV. We discuss a strategy to incorporate this
picture in the framework of superstring theory. In such a scheme 
successful predictions of GUTs can be realized while  some of
the more problematic aspects of grand unification might be avoided.
The most promising models are expected in the
framework of the heterotic $E_8\times E_8$ string theory.
}

\section{Strings and the real world}
String theory might provide a scheme that allows a unified description of all fundamental interactions. We thus need to 
compare the properties of string theory with the phenomena of the real world. First we might ask: what does string 
theory give us? In fact, it provides us with all we need: gravity, gauge interactions and matter fields in chiral 
representations of the gauge group. But we get even more than we need, e.g. extra dimensions, supersymmetry and large 
gauge groups like $SO(32)$ or $E_8 \times E_8$. In its simplest form superstring theory does not really resemble the 
real world. We observe $d=4$ space-time dimensions instead of $d=10$, $N=0$ (or $N=1$) supersymmetry instead of $N=8$ (or $N=4$) 
and gauge groups much smaller than $E_8 \times E_8$. The world we see contains $SU(3)\times SU(2)\times U(1)$ gauge 
bosons, three families of quarks and leptons and (most probably) a Higgs boson. This might fit very well in the 
framework of superstring theory, but there is no convincing argument from string theory that this should come out like that.
Independent of string theory, theoretical 
arguments lead us beyond the $SU(3)\times SU(2) \times U(1)$ standard model. Among them is the quest for gauge 
unification (GUT) at a large scale $M_{\mbox{GUT}} ~\sim 10^{16} - 10^{17} GeV$ combined with weak scale $N=1$ 
supersymmetry to avoid the hierarchy problem. There are (indirect) experimental indications that support this point of 
view. The evolution of gauge couplings in the minimal supersymmetric extension of the standard model (MSSM) points 
towards unification at $M_{\mbox{GUT}}$. Recent observations in neutrino physics suggest the existence of heavy right 
handed Majorana neutrinos in the framework of a simple see-saw mechanism within a grand unified picture. In fact 
GUT-schemes are so appealing that they have changed our view of the world. They could certainly be embedded in a string 
theory description, but again we do not have a theoretically convincing top-down prediction for a specific GUT 
realization from first principles.

In view of this situation we suggest the following procedure. Start with an educated (bottom-up) guess of a unified 
picture and then scan the string possibilities. Within this framework we would hope to get hints from string theory and 
a possible ``improvement'' of the unified picture. Can extra dimensions and properties of compactification remove some of 
the problems of GUT model building? Are there some specific stringy explanations for some of the parameters of the 
standard model? Rather than analyzing these questions in a general framework, we would like to select models (via our 
bottom-up procedure) that are already close to our observations. For this selection procedure we propose certain rules. 
The rules come from our educated (bottom-up) guess of a unified picture. Following the rules should make sure that key 
properties of particle physics phenomenology will be included in the attempts at string model building.

\section{Five Rules}
\subsection{Rule I: Spinor of $SO(10)$}
Grand Unified Theories lead in particular to a unified description of families of quarks and leptons \cite{Langacker:1980js}.
Promising GUT 
groups are e.g. $SU(5), SO(10), E_6$, with a chiral matter description of quark-lepton families in the $\bar{5} + 10, 
16, 27$ dimensional representation, respectively. As a GUT group $SU(10)$ is clearly singled out:
\begin{itemize}
\item it incorporates all the success of Georgi-Glashow $SU(5)$ and Pati-Salam $SU(4) \times SU(2) \times SU(2)$,
\item one family of quarks and leptons fits in a single irreducible representation: the 16-dimensional spinor 
\item right handed neutrinos are naturally included,
\item exotic matter candidates are avoided.
\end{itemize}
From these facts we conclude that there is no alternative compelling structure for a unified description of a family of 
quarks and leptons other than the
\begin{itemize}
\item 16-dimensional spinor representation of $SO(10)$.
\end{itemize}  
It combines the prediction of right-handed neutrinos with the unification of Yukawa-couplings in a simple mathematical 
structure. Therefore, we propose that string model building should concentrate on schemes that include the spinor 
representation of $SO(10)$ in a low-energy description.

One might now be tempted to assume that in such a scheme the full $SO(10)$ symmetry must be realized. Such an 
assumption, however, is not necessary, and this is one of the places where the consideration of string theory with its 
extra dimensions might be useful. It leads us to the second rule.

\subsection{Rule II: Incomplete Multiplets}
While complete (16-dimensional) $SO(10)$ representations might be appropriate for fermions, this does not seem to be 
the case for gauge- and Higgs-bosons. Up to now we only have seen the gauge bosons of $SU(3) \times SU(2) \times U(1)$ 
and we will hopefully see only $SU(2)$ Higgs-doublets and not the (problematic) GUT partners (e.g. $SU(3)$ triplets). 
This is the essence of the well-known ``doublet-triplet splitting problem'' of grand unified theories.

The solution to this problem is the appearance of incomplete (split) multiplets for gauge- and Higgs-bosons. This leads 
to the following questions:
\begin{itemize}
\item where are the other states?
\item how is the GUT gauge symmetry broken?
\end{itemize}
It is exactly here that the consideration of string theory in extra dimension improves the situation compared to 
conventional GUTs. It is known since the early times of orbifold compactification \cite{Dixon:1985jw} 
of heterotic string theory that the 
doublet-triplet splitting is solved there generically \cite{Ibanez:1987sn}. In a similar way GUT gauge symmetries 
are in general not 
realized in $d=4$ but only in higher dimensions. This leads us to the concept of
\begin{itemize}
\item GUTs without GUT group:
\end{itemize}
we have ``complete'' fermion representations but incomplete gauge and Higgs multiplets. In such a scheme we can 
accomodate all the successful aspects of grand unification (as e.g. gauge- and Yukawa coupling unification, complete 
fermion multiplets) while avoiding the problematic ones (such as fast proton decay and the problems of a spontaneous 
breakdown of the GUT group). It is here that string model building can drastically improve our notion of grand 
unification. We shall later come back to a more explicit discussion of that situation.

At this moment we should stress another aspect of incomplete multiplets: the 16-dimensional spinor of $SO(10)$ could be 
an incomplete representation as well. This could happen if in $d> 4$ dimensions the gauge group is even larger than 
$SO(10)$. One version of the heterotic string provides  the gauge group $E_8 \times E_8$. Thus the 16 of $ SO(10)$ 
could be a split multiplet with respect to $E_8$ or a subgroup thereof (like $E_6$) that incorporates $SO(10)$. Thus 
there is still flexibility in explicit model building. Later we shall see, however, that some of the aspects of 
$SO(10)$ are key ingredients for successful model building.  

\subsection{Rule III: Repetition of Families}
A family of quarks and leptons fits in a 16-dimensional spinor of $SO(10)$. But we observe not only one but three 
families. So far grand unified model building has failed to supply arguments why this should be so. Attempts to 
incorporate 3 families in a single representation of a GUT group have failed to provide convincing examples. Typically 
such schemes have the unattractive feature to predict too many exotic states in addition to the 3 families.
Again, theories with extra dimensions can give a clue to family repetition. In the process of compactification of a 
higher dimensional theory to $d=4$, the appearance of a repetition of matter multiplets is quite common and it depends 
on topological and geometrical properties of the compact manifolds. In the framework of Calabi-Yau compactification of 
string theory, the number of families \cite{Candelas:1985en}
is given by the Euler number of the manifold, in orbifold-compactification one 
might find geometrical explanations for the number of families \cite{Ibanez:1986tp,Kim:1992en,Faraggi:1993pr}. 
No model has emerged yet that convincingly explains why 
the number of families is 3, but still we should be satisfied to have a mechanism that naturally leads to a simple 
repetition of a family structure, as provided by (string) theories with extra dimensions. Such a ``geometrical'' 
interpretation of the family structure might be a useful ansatz in analyzing the so-called flavour problem (such as 
e.g. the absence of flavour changing neutral currents or the pattern of quark and lepton masses). With the geometrical 
picture come additional (discrete) symmetries that might be of great phenomenological relevance. Thus the flavour 
problem (and the number of families) should find an explanation from higher dimensions. No compelling 
alternative explanation is known. 

\subsection{Rule IV: $N=1$ Supersymmetry}
Why is the weak scale so small compared to the Planck scale? This is the question connected to the hierarchy problem. 
Attempts to solve the problem have been
\begin{itemize}
\item Technicolor,
\item Supersymmetry,
\item Conformal symmetry.
\end{itemize}
All of them require new particles beyond the standard model somewhere around the weak scale. Each of the suggestions 
has its merits and weaknesses. Still here we would like to argue in favour of supersymmetry. Some arguments in this 
direction are:
\begin{itemize}
\item the observed evolution of gauge couplings in the MSSM,
\item some indirect indication for a light Higgs boson,
\item arguments for a grand desert coming from neutrino see-saw and ``absence'' of proton decay,
\item the natural incorporation of the GUT picture,
\item a compelling candidate for cold dark matter.
\end{itemize}
This makes us confident to invest some time in the construction of $N=1$ supersymmetric string derived models in $d=4$ 
to solve the hierarchy problem. There are, however, remaining problems that have to be addressed. We need an 
understanding of the mechanism of supersymmetry breakdown and the so-called $\mu$-problem. In addition proton decay via 
dimension 5 - operators has to be sufficiently suppressed. For a review see ref. [8]. 

Lately, there has been a lot of activity in models with ``large'' extra dimensions. Some people say that the hierarchy 
problem is solved by either supersymmetry or large extra dimensions. Such a statement is highly misleading. One should 
rather say: small Higgs mass or large extra dimensions. In both cases one has to discuss the stability of a small 
parameter, and supersymmetry could provide an explanation (in both cases). 

\subsection{Rule V: R-parity and other Discrete Symmetries}
Our experience with supersymmetric model building tells us that (discrete) symmetries are utterly important. We need 
then to 
\begin{itemize}
\item avoid proton decay via $d=4$ operators,
\item provide a stable particle for cold dark matter,
\item explain desired textures for Yukawa couplings,
\item address the flavour problem,
\item solve the $\mu$-problem.
\end{itemize}

One of these symmetries could be the standard R-parity of the MSSM: it forbids the $d=4$ operators to avoid proton 
decay, allows the standard Yukawa couplings and leads to cold dark matter candidate (e.g. in form of a neutralino). It 
is interesting to observe that an $SO(10)$ theory with matter fermions in the spinor representation naturally provides 
this R-parity. In fact, this reinforces our arguments in Rule I for the spinor of $SO(10)$. In addition to that, 
discrete symmetries appear generically in string model building. So there is hope that the desired symmetries might be 
found.

\section{Where does this lead us?}
\subsection{Intermediate conclusion}
The arguments given so far convince us that we should aim at the construction of $N=1$ supersymmetric models that 
contain the 16-dimensional spinor representation of $SO(10)$ to 
\begin{itemize}
\item describe a family of quarks and leptons,
\item allow a natural incorporation of R-parity.
\end{itemize}
Extra dimensions beyond $d=4$ are needed to explain
\begin{itemize}
\item incomplete (split) multiplets,
\item repetition of families.
\end{itemize}
It is obvious that in such a case the extra dimensions cannot be ``large'' (like e.g. inverse TeV scale) as this would be 
incompatible with a grand unified picture. 

\subsection{Explicit string constructions}
Attempts at the construction of (semi)realistic models from string theory are based on
\begin{itemize}
\item	heterotic $SO(32)$,
\item heterotic $E_8 \times E_8$,
\item type I $SO(32)$,
\item type II (orientifolds),
\item intersecting branes $U(N)^M$,
\item M-theory on manifolds with $G_2$-holomony,
\item heterotic M-theory $E_8 \times E_8$.
\end{itemize} 

Without a detailed discussion of these individual classes of models 
it is evident that the rules can be incorporated 
most naturally within the heterotic $E_8 \times E_8$ string theory. 
For more information on explicit model building 
see ref. [9,10,11] 
The heterotic M-theory of Horava and Witten\cite{Horava:1995qa} might 
be equally suitable (representing the heterotic $E_8 \times E_8$ theory at finite coupling), but its 
formulation as an exact fundamental theory has not yet been established. Similar remarks apply to the case of M-theory on 
manifolds with $G_2$ holomony \cite{Acharya:2001gy}, 
where we still lack an explicit example that incorporates the spinor of $SO(10)$. In any 
case renewed efforts in model constructions based on the heterotic $E_8 \times E_8$ theory should be encouraged. 
Standard approaches of orbifold or Calabi-Yau compactification deserve further investigations. 
\subsection{Some group theory speculation}
The classification of Lie algebras leads to 4 infinite series $A_n , B_n , C_n , D_n$ and certain exceptional cases 
$G_2 , F_4 , E_6, E_7, $ and $E_8$. For a textbook see [14]. 
Chirality in $d=4$ would imply that only $A_n = SU(n + 1), D_n = SO(2n)$ or $E_6$ 
are acceptable. Chiral fermions are much easier to obtain in $d= 4k + 2$ (and especially $d= 8 k + 2$) dimensions. 
String theory points towards $d=10$ where even the fundamental 248-dimensional representation of $E_8$ is chiral. As 
$E_8$ can be easily realized in string theory one might speculate that a finite $E_n$ series might
be particularly suited for a description of grand unification 
in the real world. The series starts with $E_8$: the other members are obtained by removing appropriate nodes from the 
Dynkin diagram of $E_8$.\\
\begin{figure}[ht]
\centerline{\epsfxsize=2in\epsfbox{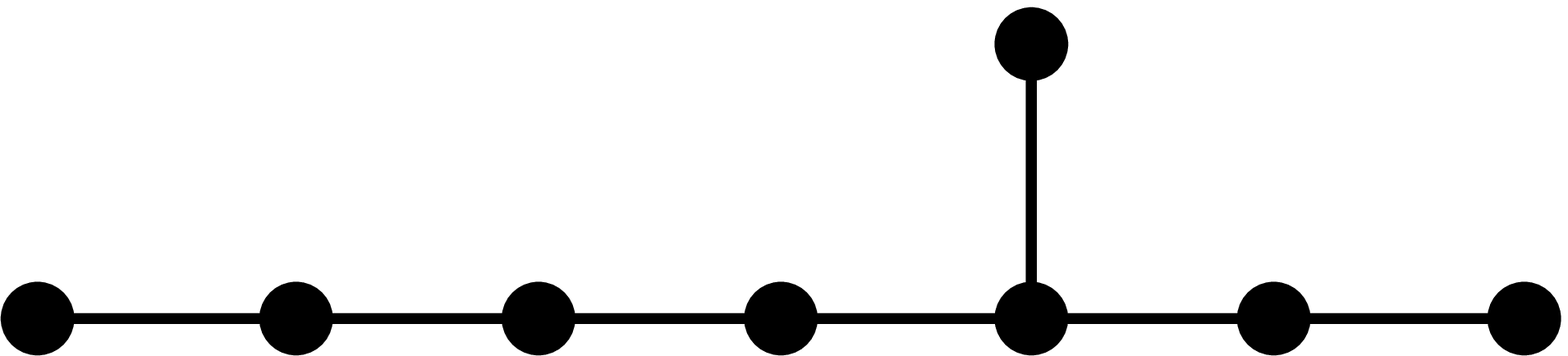}}
\end{figure}\\
In a first step this leads to $E_7$,\\
\begin{figure}[ht]
\centerline{\epsfxsize=5cm   
\epsfbox{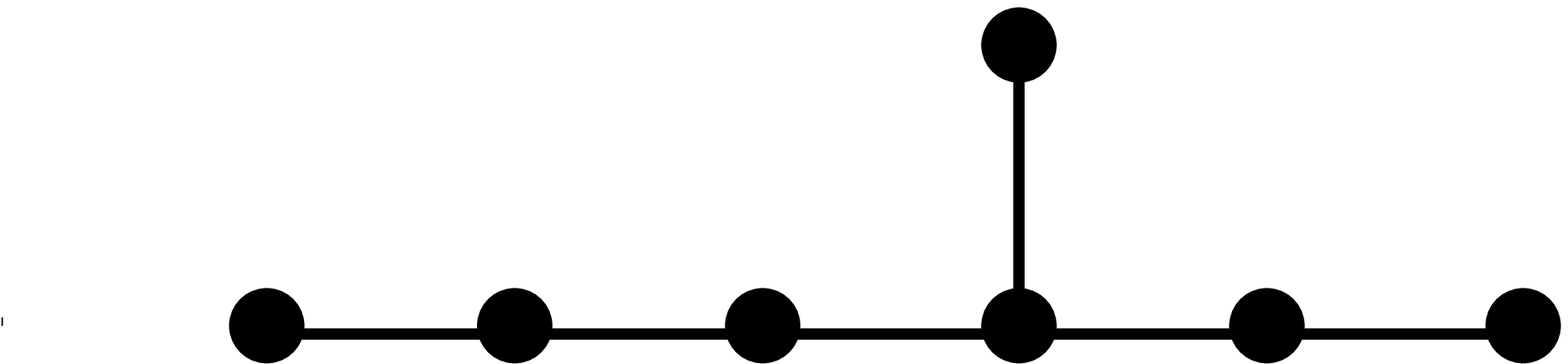}}
\end{figure}\\
then to $E_6$,
\begin{figure}[ht]
\centerline{\epsfxsize=5cm   
\epsfbox{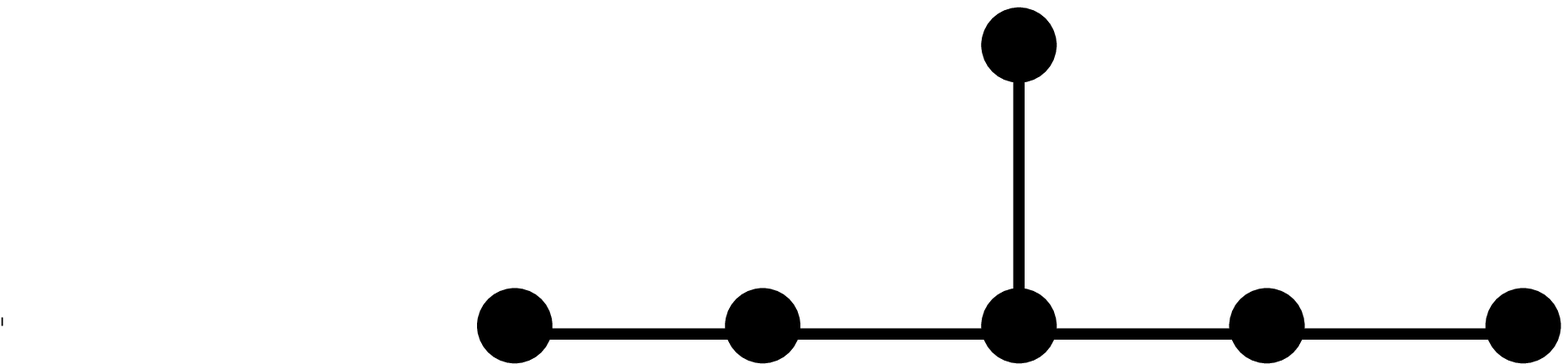}}
\end{figure}\\
and further to $E_5 \equiv D_5 = SO(10)$. 
\\
\begin{figure}[ht]
\centerline{\epsfxsize=5cm   
\epsfbox{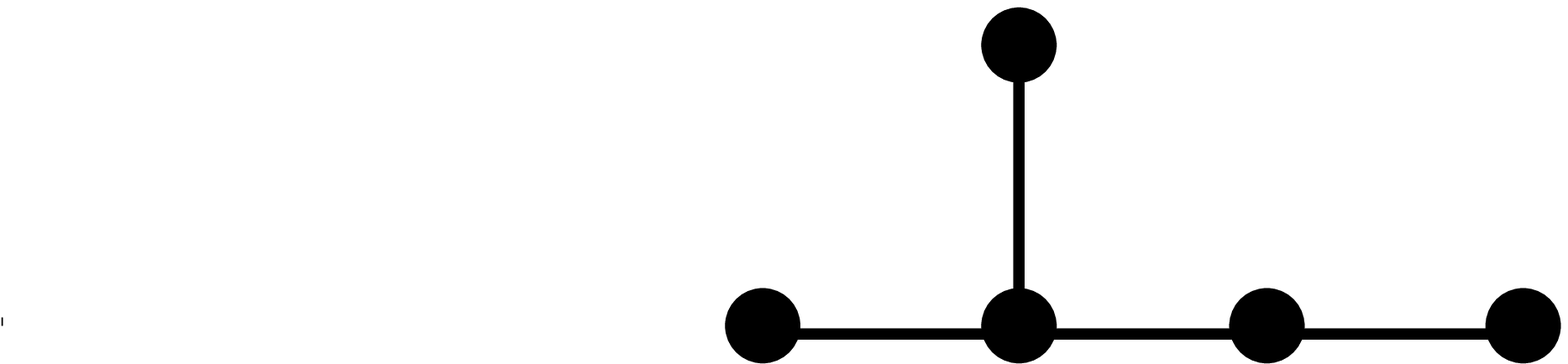}}
\end{figure}\\
$E_4$ coincides with $A_4 = SU(5)$\\
\begin{figure}[ht]
\centerline{\hskip3.1cm\epsfxsize=0.7in\epsfbox{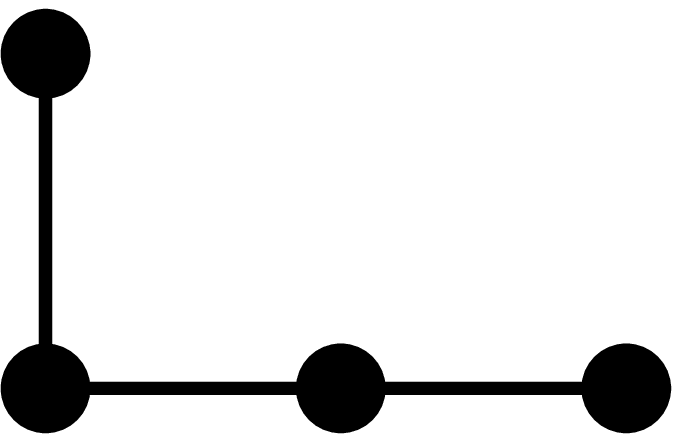}}
\end{figure}\\
and finally $E_3 \equiv SU(3) \times SU(2) \times (U(1))$.\\
\begin{figure}[ht]
\centerline{\hskip3.1cm\epsfxsize=0.7in\epsfbox{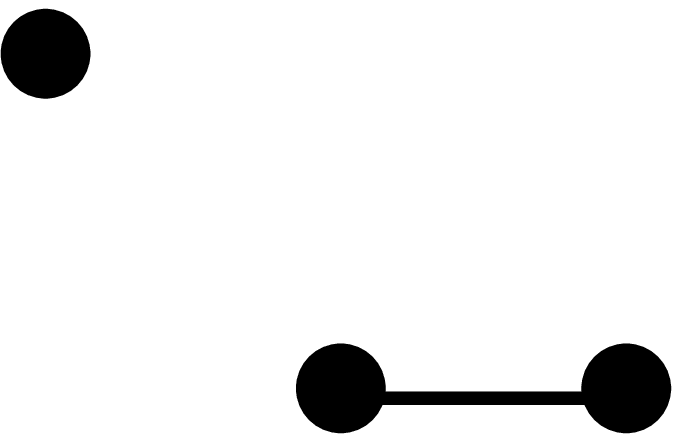}} 
\end{figure}
\\

All the successful gauge groups of $d=4$ grand unification like $E_6, SO(10), SU(5)$ and even the standard model are 
encoded in this $E_n$ series, and the connection with string theories in $d > 4$ leads (via $E_7$) to $E_8$. Many 
properties of effective $d=4$ unified theories might be attributed to the presence of a higher dimensional $E_n$ group.

\section{The memory of $E_5 = SO(10)$}
Our bottom-up approach showed that $SO(10)$ symmetry is a very efficient way to describe a grand unified extension of 
the standard model and this found its reflection in the formulation of the rules given in Section 2. In explicit model 
building one might, of course, ask the question what could go wrong if one violates the rules (as is done in many attempts 
of string model building). The first thing that could go wrong is the identification of $U(1)_Y$-hypercharge and the 
question of gauge unification. $SO(10)$ will give you the (more or less) correct value of the weak mixing angle $\sin^2 
\theta_W$, while in other cases more often than not $U(1)_Y$ is a combination of many $U(1)'s$ and ``unconventional'' 
values of $\sin^2 \theta_W$ are obtained. Apart from this,
$SO(10)$ with a spinor can very well accomodate the fermion spectrum, including 
right handed neutrinos. In alternative constructions, the right handed neutrinos have to be identified among a host of 
singlets. Yukawa unification (at least for the third family) would follow from $SO(10)$. In addition the inherited 
R-parity would forbid unwanted terms that could lead to proton decay via dimension 4 operators. In models based on 
$SO(32)$ the situation is usually the opposite: forbidden top-quark Yukawa coupling and allowed $d=4$ 
proton decay operators. More model building and new symmetries would then be needed to assure the stability of the 
proton. Of course, if you start with $SO(10)$ and the spinor it is not guaranteed that you obtain a successful model, 
but at least you avoid the obvious things that might go wrong.

In fact, we have seen that not all of the aspects of $SO(10)$ grand unification might be desirable and we would rather 
argue for split multiplets for gauge- and Higgs-bosons. Thus $SO(10)$ should be realized in $d > 4$ and broken in 
$d=4$. Many of the properties of the $d=4$ low energy effective theory would then depend on
\begin{itemize}
\item properties of the underlying $SO(10)$ in higher dimensions (the memory of $SO(10)$),
\item geometrical aspects of compactification.
\end{itemize}  
The geometrical aspects would be relevant for the number of families and thus the potential family symmetries. The 
interplay of ``$SO(10)$ memory'' and geometrical structure would allow a description of many observable facts. Let us  
just discuss two examples. The third family might be exposed to a higher degree of gauge symmetry 
(e.g. being localized 
on a so-called $SO(10)$ brane: a localized region in extra dimensions, where the full $SO(10)$ symmetry can be felt) 
than the other two families. This would explain why the Yukawa couplings of the third family 
respect $SO(10)$ mass relations but not the other two families. With respect to proton decay too much gauge symmetry 
might be dangerous. Thus we find a suppression of proton decay if the first family does not feel the full $SO(10)$ 
symmetry.
\section{Conclusions}
At the moment, string model building needs some bottom-up input. An educated guess would then suggest constructions 
addressing the following basic questions:
\begin{itemize}
\item how to get a spinor of $SO(10)$, 
\item ``small'' gauge group in $d=4$,
\item three families of quarks and leptons,
\item split multiplets for Higgs bosons,
\item gauge unification $(\sin^2\theta_W)$,
\item R-parity and proton decay,
\item (partial) Yukawa unification,
\item the flavour problem and (discrete) family symmetries.
\end{itemize} 
Such a procedure should allow the identification if $N=1$ supersymmetric
models that could serve as a starting point for a realistic generalization 
of the standard model of particle physics. 

Following the early attempts\cite{Ibanez:1987pj} in the 80's, 
some standard-like models\cite{Faraggi:1991jr,Cleaver:1998sa}
have been found in the 90's. In the last years, promising models 
have been analyzed in the framework of orbifold 
compactifications\cite{Choi:2003ag,Kobayashi:2004ud,Forste:2004ie,Kobayashi:2004ya},
the free fermionic formulation\cite{Faraggi:2004rq,Donagi:2004ht} and
Calabi-Yau compactification\cite{Braun:2004xv}. Ultimately one might like
to obtain a connection to succesful field theory orbifold 
schemes\cite{Asaka:2003iy}.

Remaining questions, not 
addressed in this talk, would be
an understanding of supersymmtery breaking and moduli stabilization in 
this class of models. These are very difficult problems that need new 
conceptual developments in string theory. 
Towards the question of moduli stabilization some progress has recently been
made in the framework of flux
compactifications.
So there is hope that
finally we shall be able to meet the real world via a string theory 
description.

\section{Acknowledgements}
This is a written up version of talks given at the conference on ``String
Phenomenology 2004, Ann Arbor, and ``Pascos04-NathFest 
Symposium'', Boston, August 2004. 
This work was partially supported by the European community's Marie Curie
programs MRTN-CT-2004-503369 ``Quest for Unification'' and
MRTN-CT-2004-005104 ``Forces Universe''.

%
%
%
%


\begin{thebibliography}{0}

\bibitem{Langacker:1980js}
P.~Langacker,
Phys.\ Rept.\  {\bf 72} (1981) 185.

\bibitem{Dixon:1985jw}
L.~J.~Dixon et al.,
Nucl.\ Phys.\ B {\bf 261} (1985) 678.

\bibitem{Ibanez:1987sn}
L.~E.~Ibanez et al.,
Phys.\ Lett.\ B {\bf 191} (1987) 282.

\bibitem{Candelas:1985en}
P.~Candelas, G.~T.~Horowitz, A.~Strominger and E.~Witten,
Nucl.\ Phys.\ B {\bf 258} (1985) 46.

\bibitem{Ibanez:1986tp}
L.~E.~Ibanez, H.~P.~Nilles and F.~Quevedo,
Phys.\ Lett.\ B {\bf 187} (1987) 25.

\bibitem{Kim:1992en}
H.~B.~Kim and J.~E.~Kim,
Phys.\ Lett.\ B {\bf 300} (1993) 343
[arXiv:hep-ph/9212311].

\bibitem{Faraggi:1993pr}
A.~E.~Faraggi,
Phys.\ Lett.\ B {\bf 326} (1994) 62
[arXiv:hep-ph/9311312].

\bibitem{Nilles:1983ge}
H.~P.~Nilles,
Phys.\ Rept.\  {\bf 110} (1984) 1.

\bibitem{Quevedo:1996sv}
F.~Quevedo,
arXiv:hep-th/9603074.

\bibitem{Uranga:2003nq}
A.~M.~Uranga,
Fortsch.\ Phys.\  {\bf 51} (2003) 879.

\bibitem{Faraggi:1992fa}
A.~E.~Faraggi,
Nucl.\ Phys.\ B {\bf 387} (1992) 239
[arXiv:hep-th/9208024].

\bibitem{Horava:1995qa}
P.~Horava and E.~Witten,
Nucl.\ Phys.\ B {\bf 460} (1996) 506
[arXiv:hep-th/9510209].

\bibitem{Acharya:2001gy}
B.~Acharya and E.~Witten,
arXiv:hep-th/0109152.

\bibitem{georgi}
H. Georgi, Lie Algebras in Particle Physics, Frontiers in Physics,
Perseus Books, 1999

\bibitem{Ibanez:1987pj}
L.~E.~Ibanez, J.~Mas, H.~P.~Nilles and F.~Quevedo,
Nucl.\ Phys.\ B {\bf 301}, 157 (1988).


\bibitem{Faraggi:1991jr}
A.~E.~Faraggi,
Phys.\ Lett.\ B {\bf 278} (1992) 131.

\bibitem{Cleaver:1998sa}
G.~B.~Cleaver, A.~E.~Faraggi and D.~V.~Nanopoulos,
Phys.\ Lett.\ B {\bf 455} (1999) 135
[arXiv:hep-ph/9811427].

\bibitem{Choi:2003ag}
K.~S.~Choi and J.~E.~Kim,
Phys.\ Lett.\ B {\bf 567} (2003) 87
[arXiv:hep-ph/0305002].

\bibitem{Kobayashi:2004ud}
T.~Kobayashi, S.~Raby and R.~J.~Zhang,
arXiv:hep-ph/0403065.

\bibitem{Forste:2004ie}
S.~Forste, H.~P.~Nilles, P.~K.~S.~Vaudrevange and A.~Wingerter,
arXiv:hep-th/0406208.

\bibitem{Kobayashi:2004ya}
T.~Kobayashi, S.~Raby and R.~J.~Zhang,
arXiv:hep-ph/0409098.

\bibitem{Faraggi:2004rq}
A.~E.~Faraggi, C.~Kounnas, S.~E.~M.~Nooij and J.~Rizos,
Nucl.\ Phys.\ B {\bf 695} (2004) 41
[arXiv:hep-th/0403058].

\bibitem{Donagi:2004ht}
R.~Donagi and A.~E.~Faraggi,
Nucl.\ Phys.\ B {\bf 694} (2004) 187
[arXiv:hep-th/0403272].

\bibitem{Braun:2004xv}
V.~Braun, B.~A.~Ovrut, T.~Pantev and R.~Reinbacher,
arXiv:hep-th/0410055.

\bibitem{Asaka:2003iy}
T.~Asaka, W.~Buchmuller and L.~Covi,
Phys.\ Lett.\ B {\bf 563} (2003) 209
[arXiv:hep-ph/0304142].



































\end{thebibliography}
\end{document}